\documentclass{mn2e}
\usepackage{times}
\input{psfig.sty}

\begin{document}

\title[X-ray jets of SS~433]
{Rapid variability of the arcsec-scale X-ray jets of SS~433}
\author[S. Migliari et al.]
{S. Migliari$^1$\thanks{migliari@science.uva.nl}, R. P. Fender$^1,^2$,
K. M. Blundell$^3$, M. M\'endez$^4$, M. van der Klis$^1$\\
\\
$^1$ Astronomical Institute `Anton Pannekoek', University of Amsterdam, and Center for High Energy Astrophysics, Kruislaan 403, \\ 
1098 SJ, Amsterdam, The Netherlands.\\
$^2$ Department of Physics and Astronomy, University of Southampton, Hampshire
SO17 1BJ, UK\\
$^3$ Astrophysics, University of Oxford, Denys Wilkinson Building, Keble Road,
Oxford OX1 3RH, UK\\ 
$^4$SRON, National Institute for Space Research, 3584 CA, Utrecht, The
Netherlands\\  
}

\maketitle

\begin{abstract}

We present the X-ray images of all the available Chandra observations of the
galactic jet source SS~433. We have studied the morphology of the X-ray images
and inspected the evolution of the arcsec X-ray jets, recently found to be
manifestations of in situ reheating of the relativistic gas downstream in the
jets. The Chandra images reveal that the arcsec X-ray jets are not steady long
term structures; the structure varies, indicating that the reheating processes
have no preference for a particular precession phase or distance from the
binary core. Three observations made within about five days in May 2001, and a
60 ks observation made in July 2003 show that the variability of the jets can
be very rapid, from timescales of days to (possibly) hours.
The three May 2001 images show two resolved knots in the east jet
getting brighter one after the other, suggesting that a common phenomenon
might be at the origin of the sequential reheatings of the knots. We discuss
possible scenarios and propose a model to interpret these brightenings in
terms of a propagating shock wave, revealing a second, faster outflow in the
jet.

\end{abstract}

\begin{keywords}

binaries: close -- stars: binaries : individual:  SS 433 -- 
ISM: jets and outflows radio continuum: stars 

\end{keywords}

\section{Introduction}

SS~433 is one of the most powerful and most studied jet sources in our
Galaxy. It is a high-mass X-ray binary system located at a distance of about
5~kpc [e.g. $4.85\pm0.2$ in Vermeulen et al. (1993a); $4.61\pm0.35$ in
Stirling et al. 2002; the most recent and likely estimate is $5.5\pm0.2$
in Blundell \& Bowler (2004)]. The nature of the accreting compact object,
either black hole (BH) or neutron star (NS), is still controversial. The most
remarkable feature of the source is the presence of bipolar, precessing,
mildly-relativistic jets. These jets are observed at wavelengths from radio to
X-rays and have been widely studied over about 30 years.

The geometry of the jets is well described to first order by the so-called
`kinematic model' (Abel \& Margon 1979; Milgrom 1979; Hjellming \& Johnston
1981; see Eikenberry et al. 2001 and Stirling et al. 2002 for an update): the
system is ejecting matter in the form of two anti-parallel narrow jets, with
an opening angle less than $5^{\circ}$. These jets precess in a $21^{\circ}$
half-opening angle cone with a period of about 162.4 days (e.g. Eikenberry et
al. 2001). The individual jet components move ballistically away from the
system. The angle between the precession axis and the line of sight is about
$78^{\circ}$, and the projection of the jets onto the plane of the sky results
in a twisted trace. A velocity of $\sim0.26c$ has been inferred from
Doppler-shifted optical emission lines in the thermal component of the
jets. The velocity of the jets seems to be stable, although a $\la15$ per cent
symmetrical scattering in the Doppler-shifts of the optical lines (see
Eikenberry et al. 2001) can be interpreted as an actual scatter of the
velocity of the optical emitting jet component, i.e. at least up to distances
of $\sim10^{15}$~cm from the binary core, where the optical emission lines
originate (e.g. Shaham 1981).

SS~433 is the only known galactic relativistic jet source to reveal the
presence of baryonic matter in the jets, in the form of emission lines (both
in optical and X-rays; e.g. Margon et al. 1979; Kotani et al. 1994) from
thermal gas. The model developed and generally accepted to describe the
emission of the thermal component of the jets is the `adiabatic cooling model'
(Brinkmann et al. 1991; Kotani et al. 1996; see also Marshall et
al. 2002). The temperature and the density of the jet gas decrease with
increasing distance from the binary core. Close to the binary core, at
distances less than $10^{11}-10^{12}$~cm, the jet gas is at a temperature of
about $10^{8}$~K (consistent with the temperatures of the inner regions of the
accretion disc), and emits in the X-ray band. Subsequently the gas in the jets
cools; at distances of about $10^{15}$~cm from the core the temperature is of
the order of $10^{4}$~K and the gas emits optical radiation. Based on the
adiabatic cooling model, beyond this point the gas should be too cool to
thermally emit optical or X-ray radiation.  An alternative model has been
proposed in which a high temperature plasma emitting in X-rays coexists with
clumps of cold matter emitting optical lines (Bodo et al. 1986; Brinkmann et
al. 1988).
Non-thermal (synchrotron) radio emission from the jets is observed at
distances of $10^{15}-10^{17}$~cm from the binary core (e.g. Hjellming \&
Johnston 1981).
\begin{figure*}
\centering
\begin{tabular}{c}
\psfig{figure=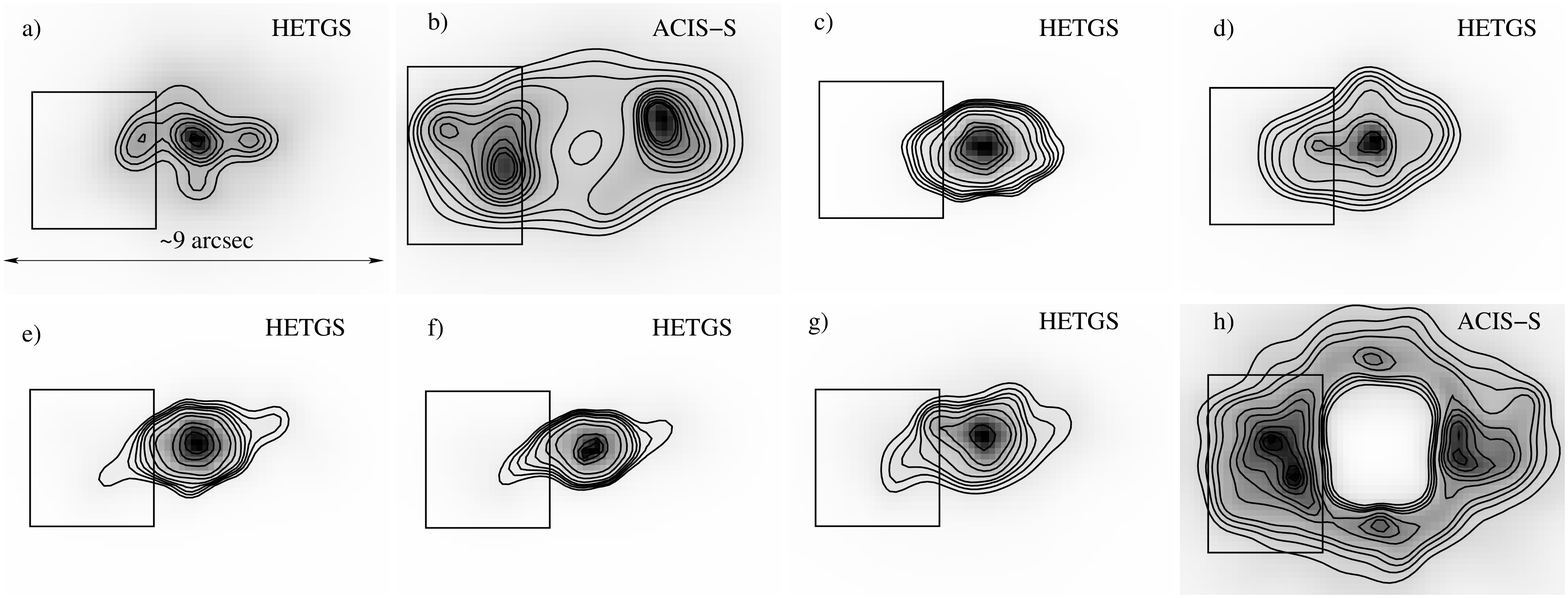,width=17cm,angle=0}\\
\psfig{figure=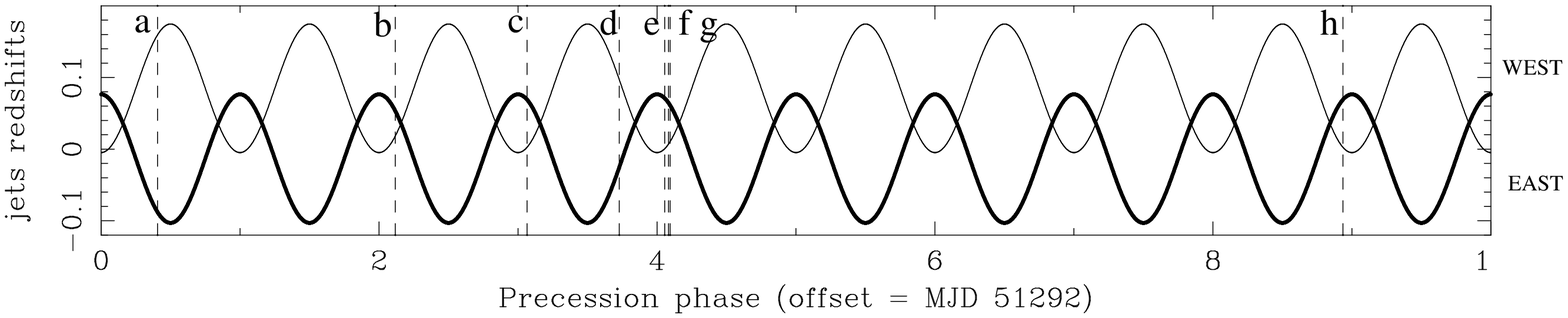,width=16cm,height=2.5cm,angle=0}\\
\end{tabular}
\caption{{\em Upper panels}: smoothed zeroth order HETGS and ACIS-S images of
all the Chandra observations of SS~433: see Table~1 for details on each
observation. The contour levels of HETGS images are normalised to the exposure
time of {\em g} which has contours at 2.5, 3, 3.5, 4, 5, 7, 10, 20, 27, 34
counts per rebinned and smoothed pixel. The contour levels of ACIS-S
observations are normalised to the exposure time of observation {\em h} which
has contours at 4, 5, 6, 7, 10, 12, 14, 17, 20, 22, 24. The color scales of
the pixel images are normalised to the counts at the peak of each image. The
regions superimposed to the images have been used to calculate the count rates
in the east arcsec scale jets. The regions have the same area and the same
physical coordinates (RA and Dec.) for all the images taken with the same
instrument, but different between HETGS and ACIS-S observations. {\em Lower
panel}: redshifts of the east (thick line) and west (thin line) jets at the
base of the jet vs. precession phase. Ticks in the absissa mark a complete
precessing cycle, i.e. every 162.4 days. The precession phase is calculated
based on ephemeris in Stirling et al. (2002):
$\phi=[(obs(JD)-2440000.5)-8615.5]/162.4$. We show with vertical dotted lines
the (core) precession phases at the time the Chandra observations have been
taken.}
\end{figure*}

Recent observations, however, appear to contradict, at least partially, the
predictions of the adiabatic cooling model. In an observation taken with the
Chandra high energy transmission grating spectrometer (HETGS), Marshall et
al. (2002) first discovered arcsec scale X-ray jets in SS~433. Further
spatially resolved X-ray spectral analysis of the arcsec jets, using a Chandra
advanced charge-coupled device imaging spectrometer (ACIS)-S observation,
revealed highly-ionised Doppler shifted iron emission lines at distances of
about $10^{17}$~cm from the binary core (Migliari, Fender \& M\'endez
2002). This indicated that in situ re-heating of atoms takes place in the jet
flow, which is still moving with (mildly-) relativistic velocity more than 100
days after launch from the binary core. Migliari et al. (2002) also attempted
to explain this re-heating by means of shocks formed due to collisions of
different blobs launched with a slightly different velocity at the same
precession phase, i.e. a faster blob which catches up a slower blob ejected
later. 

In this paper we present all the currently available Chandra observations of
SS~433. We report on the morphology and evolution of the extended X-ray
emission regions and present evidence for long-term and daily variability of
the arcsec X-ray jets, discussing possible physical scenarios.

\section{Observations and data analysis}

\begin{table*}
\centering
\caption{Observation (refering to Fig.~1); Chandra instrument; date of
observation; exposure time; orbital phase of the binary from Dolan et
al. (1997): [$obs(JD)-2450023.69$]$/13.08$; precession phase of the jet based
on Stirling et al. (2002; see Fig.~1); background subtracted count rates of
the core as estimated from the fit of the energy spectra for the HETGS
observations and from the readout streak for the ACIS observations (see
\S~2.1.2 for details); background subtracted count rates of the east jets of
all the Chandra observations of SS~433 estimated in the regions shown in the
upper panel of Fig.~1; net count rates in the regions shown in the upper panel
of Fig.~1, after the subtraction of the contribution of the core (see \S~2.2).
Uncertenties are $1\sigma$ statistical errors.}
\begin{tabular}{l l l l l l l l l}
\hline
Obs. & Instrument & Date & Exp.  & Orbit. & Prec. & core & east jet
measured&$\Delta$east jet\\
     &    &   & (ks)   & phase  & phase & (counts/s) & (counts/s)& (counts/s)\\
\hline
a$^{1}$ & HETGS  & 1999 Sept. 23 (MJD 51444) & 28.6 & 0.664 &0.420 &$1.442$ & 
$0.0298\pm0.0010$ &$0.0036\pm0.0014$\\
c       & HETGS  & 2000 Nov. 28 (MJD 51876) & 22.7 & 0.618 &0.080 & $0.557$ &
$0.0282\pm0.0011$ &$0.0183\pm0.0013$\\ 
d       & HETGS  & 2001 March 16 (MJD 51984)& 23.4 & 0.980&0.748 & $1.876$ &
$0.0598\pm0.0016$ &$0.0246\pm0.0021$\\ 
e       & HETGS  & 2001 May 8   (MJD 52037) & 19.6 &0.005 &0.072 & $0.363$ & $0.0295\pm0.0012$ &$0.0234\pm0.0013$\\ 
f       & HETGS  & 2001 May 10  (MJD 52039) & 18.5 &0.147 &0.083 & $0.234$ & $0.0293\pm0.0013$ &$0.0254\pm0.0014$\\ 
g$^{2}$ & HETGS  & 2001 May 12  (MJD 52041) & 19.7 & 0.300&0.096 & $0.846$ &
$0.0391\pm0.0014$ & $0.0238\pm0.0017$\\ 
\hline
b$^{3}$ & ACIS-S & 2000 June 27 (MJD 51722) &  9.7 & 0.890 &0.130 &$2.3$ &
$0.0803\pm0.0028$ & $0.0588\pm0.0032$\\ 
h       & ACIS-S & 2003 July 10  (MJD 52830)& 58.1 & 0.641 &0.956 & $8.8$ &
$0.0918\pm0.0013$ & $0.0132\pm0.0018$\\
\hline

\end{tabular}
\flushleft
For a more detailed analysis see:   
{\bf 1:} Marshall, Canizares \& Schulz (2002);  
{\bf 2:} Namiki et al. (2003);
{\bf 3:} Migliari, Fender \& M\'endez (2002).

\end{table*}

We have inspected the images of all the Chandra observations of SS~433 to
date: six HETGS and two ACIS-S obervations. We show all the images in
Fig.~1. We have analysed the observations with the standard tools in CIAO v
3.1. The images shown in Fig.~1 have been rebinned to one sixteenth of the
original pixel size and smoothed with the tool {\em csmooth}, using a minimal
significance signal to noise ratio of 3 and a Gaussian convolution kernel. The
observations span about four years, from 1999 September 23rd (Fig.~1, $a$) to
2003 July 10th (Fig.~1, $h$), with exposure times from 9.7~ks to 58.1~ks (see
Table~1). In the lower panel of Fig.~1 we show the redshifts due to the
precession of the east and the west jets as a function of time. The
dotted-lines indicate the days the observations have been taken. Note that
these redshifts refer to the matter launched at the base of the jets, close to
the binary core. The outer parts of the arcsec-scale jet analysed in this work
(see below) and indicated by the regions in the upper panels of Fig.~1 have a
different precession phase (and thus redshift) than the one indicated by the
dotted line in the lower panel. Matter in the jet at those distances from the
core (about $10^{17}$~cm) has been launched 100-200 days before the
observation was taken, and therefore with a precession phase corresponding to
1-1.5 precession cycles before (using the ephemeris in Stirling et
al. 2002). For each observation we have estimated the count rate of the core
(see \S~2.1.1 and \S~2.2.2) and extracted the count rate in a region of the
east jet (Fig.~1, upper panels).
In the HETGS observations the regions have the same area and are at the same
distance (centred at about $2\arcsec$) to the four central brightest rebinned
pixels in the smoothed image. In the two ACIS-S obervations we fixed the
coordinates (RA and Dec.) of the regions and chose a slightly bigger area in
order to cover the knots observed in the jets. All the quantitative analysis
in this work, such as the extractions of counts and count rates, has been done
using the original images (i.e. not rebinned and not smoothed) in the
0.5-10~keV energy range. In Table~1 we report the background subtracted count
rates in the core and in the east jet regions, with $1\sigma$ statistical
errors, of all the observations. The count rates of the ACIS observations are
much higher than the HETGS observations mostly because of the higher
sensitivity of the instrument by about a factor of four.

\subsection{Estimate of the core count rates}

\begin{table*}
\centering
\caption{Observation (refering to Fig.~1); equivalent neutral hydrogen column
density ${\rm N}_{H}$; power law photon index $\Gamma$; normalisation of the
power law N$_{PL}$; best fit $\chi^2/d.o.f.$; unabsorbed flux F in the range
2-10 keV (except observation $g$). The uncertainties are $1\sigma$ statistical
errors.}

\begin{tabular}{l l l l l l}
\hline
Obs. & ${\rm N}_{H}$ ($\times10^{22}$)  & $\Gamma$ & N$_{PL}$($\times10^{-2}$) & $\chi^2/d.o.f.$& F$_{2-10}$ ($\times10^{-11}$)\\
     & (cm$^{2}$) & & (ph cm$^{-2}$ s$^{-1}$)      & &
(erg~cm$^{-2}$~s$^{-1}$)\\ 
\hline
$a^{1}$ & $0.95$     & $1.35$ & $1.5$ &  -  &  $10.91$\\
$c$       & $1$ ($fixed$)  & $1.87\pm0.05$ &$1.03\pm0.04$& $1486/1148$ &$3.0\pm0.2$\\ 
$d$       & $1$ ($ fixed$)  & $1.47\pm0.02$ &$2.07\pm0.04$& $978/988$ &$13.36\pm0.2$\\ 
$e$       & $1$ ($ fixed$)  & $1.91\pm0.08$ &$0.68\pm0.04$& $1068/938$&$1.9\pm0.1$\\ 
$f$       & $1$ ($ fixed$)  & $1.94\pm0.11$ &$0.46\pm0.04$& $116/230$& $1.2\pm0.1$\\ 
$g^{2}$ & $1.31\pm0.06$  & $1.40\pm0.04$ &$1.18\pm0.07$ & $2312/1964$ & $8.7^{3}$\\ 
\hline

\end{tabular}
\flushleft
{\bf 1:} From Marshall, Canizares \& Schulz (2002)  
{\bf 2:} From Namiki et al. (2003) 
{\bf 3:} The flux has been estimated in the range 1-10~keV
\end{table*}

\subsubsection{HETGS observations}

\begin{figure}
\centering{\psfig{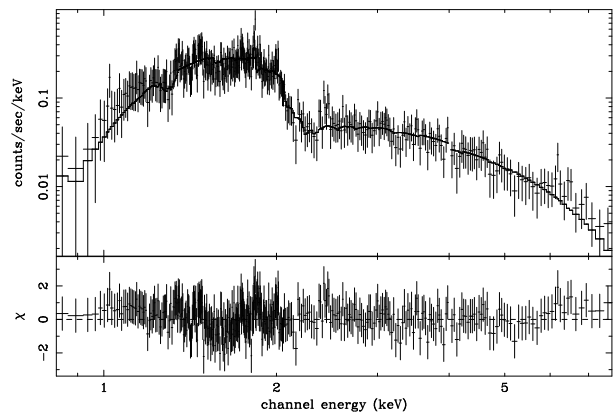}}
\caption{Energy spectrum of observation $c$ with the  residuals with respect
to the fit with an absorbed power law model. See \S~2.1.1 for details.}
\end{figure}
All the Chandra images shown in Fig.~1 are affected by pile-up. Therefore, in
the case of the HETGS observations we have estimated the count rates from the
unresolved core of SS~433 (which dominates the X-ray emission from the source)
by analysing the dispersed energy spectra, which are not affected by
pile-up. We have extracted the first order ($\pm1$) MEG and HEG spectra
following the standard procedures, using CIAO v. 3.1. We have analysed the
combined MEG and HEG spectra in the range 0.8-8.0 keV.  As in this work we are
mainly interested in the estimate of the total flux in each spectrum, which
is dominated by the broad continuum, we did not concentrate our attention on
the weaker emission lines. We have rebinned the data in order to have a
minimum of 20 counts per energy bin. The best-fit model for the continuum in
all the HETGS observations is a power law corrected for photoelectric
absorption. We have added narrow Gaussian emission lines to this model for the
spectrum of observation $d$ to fit the strongest lines with residuals larger
than $4\sigma$, namely at about $1.85$~keV, $1.98$~keV, $6.30$~keV and
$6.59$~keV. Since weaker emission lines below a few keV may still contribute
to the actual value we measure of the absorption column density,
we have fixed it to a reasonable value of N$_{H}=10^{22}$~cm$^{-2}$, which is
consistent with that found by Marshall et al. (2002) and Namiki et
al. (2003). In Table~2 we report the best-fit parameters of the continuum of
each spectrum and the unabsorbed flux in the 2-10 keV range. For observations
$a$ and $g$ the values are those in Marshall et al. (2002) and Namiki et
al. (2003), respectively. In Fig.~2 we show, as an example, the energy
spectrum with the residuals with respect to the model of observations
$c$. Although the parameters of the power law component shown in Table~2 may
be considered as approximate estimates of the physical parameters of the
source, comparing the fluxes we obtained with our fits and the fluxes derived
from the more detailed analysis in Marshall et al. (2002) and Namiki et
al. (2003), we find that they are consistent within less than 10 per cent (8
and 5 per cent, respectively, for the two observations). We have used the 2-10
keV unabsorbed fluxes of the spectra to estimate with PIMMS the 0.5-10~keV
count rates we expect from the unresolved core of the source in the Chandra
HETGS images, dividing by a factor of four the expected count rates in the
ACIS-S. The core count rates are reported in Table~1.

\subsubsection{ACIS observations}

The ACIS observations are also affected by pile-up, but there is no dispersed
spectrum. Therefore, we have estimated the 0.5--10~keV count rate of the core
by analysing the photons in the readout streak, which should be unaffected by
pile-up. We have extracted the background subtracted counts in a rectangular
region with $n\times m$ pixels covering the readout streak (where we chose
$m=20$ in a direction perpendicular to the streak and $n=280$ as the number of
row-pixels in the direction of the streak) and divided these counts by the
effective exposure time for the readout streak data, calculated as
follows. The time to read out one row of pixels in a direction perpendicular
to the readout streak is 40~$\mu$s. The effective exposure time per frame for
the streak region is then $n\times 4\times10^{-5}$~s. The number of frames of
an observation is the ratio between the actual exposure time and the frame
time (which is 3.2~s for all the SS~433 observations). The core count rates of
the ACIS observations are shown in Table~1.

\subsection{Estimate of the contribution of the core to the measured count
rates in the east jet}

\begin{figure}
\centering{\psfig{figure=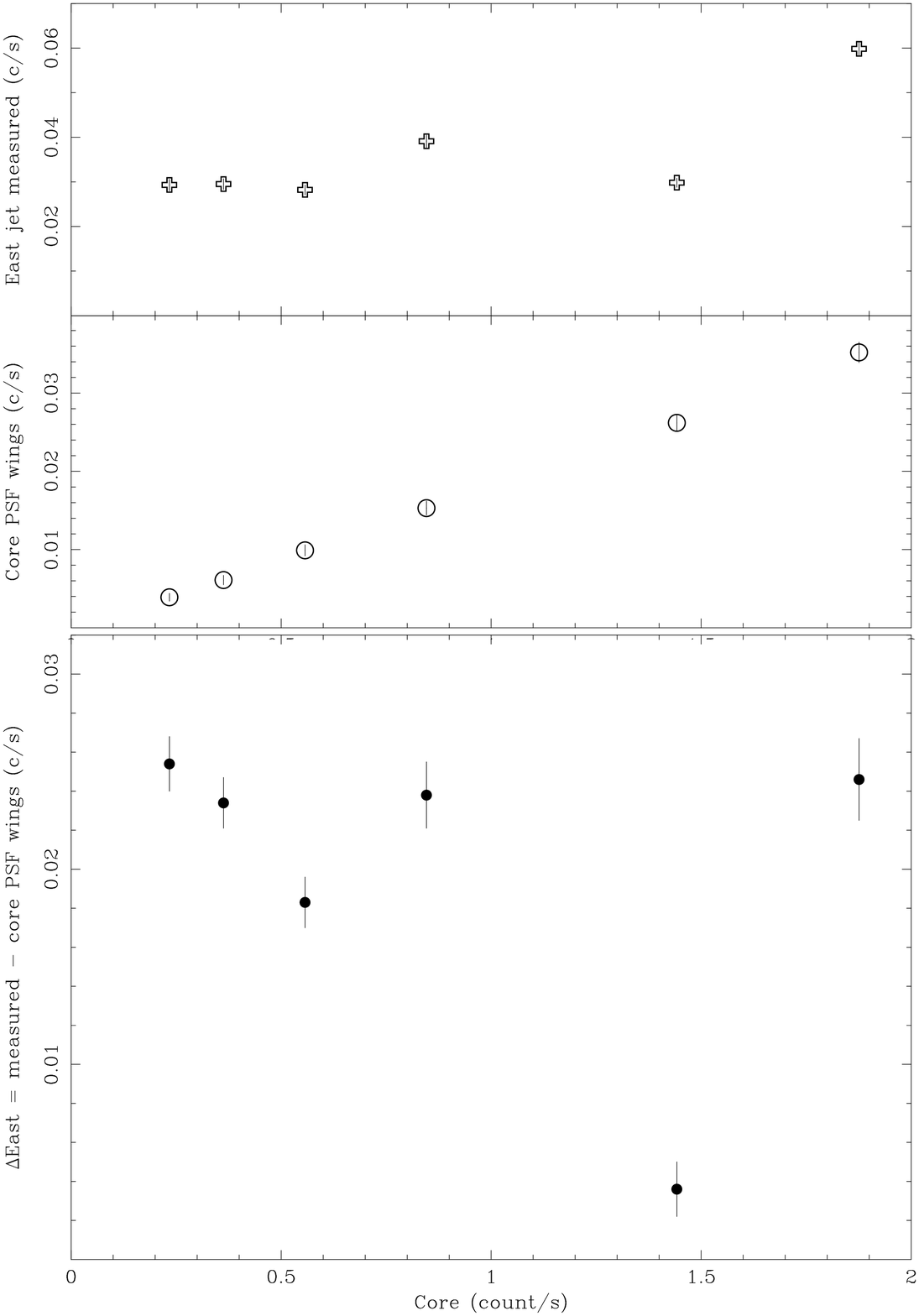,width=8cm,angle=0}}
\caption{Count rates in the east jet regions (see Fig.~1) of the HETGS
observations as measured in the image (upper panel), due to the contribution
of the bright core (middle panel) and the residuals (lower panel), as a
function of the total count rates of the core. The error bars are $1\sigma$
statistical errors.}
\end{figure}

Using the actual count rate we expect from the core of SS433 in each
observation, we have calculated the contribution of the core in the regions
where we have estimated the count rates of the east jet. For each observation
we have created the 3~keV PSF in the core position and normalised it to the
total counts in the core. We have estimated the count rate of the normalised
PSF in the regions shown in Fig.~1. The contribution of the core to the count
rates in these regions (which is $\sim2$\% of the total core count rate),
together with the total count rates we measured from the image, and the
residuals, are shown as a function of the total count rates of the core, for
the HETGS observations, in Fig.~3 (see also Table~1). It can be seen that the
wings of the core PSF may contribute up to 50\% of the flux measured in the
eastern jet. However, a constant model fit to the residuals, corresponding to
a steady eastern jet, can be rejected at the $>99.9$\% level
($\chi^2/d.o.f.=169/5$). The rejection level remains at $>99.9$\%
($\chi^2/d.o.f.=16.7/4$) even if the most strongly deviating point, that of
observation $a$ (1.44 core count/sec) is removed. This establishes that the
eastern jet is significantly variable in a way which cannot be explained
simply by the PSF of the variable core.

\section{Results}
\begin{figure*}
\centering
\begin{tabular}{c}
\psfig{figure=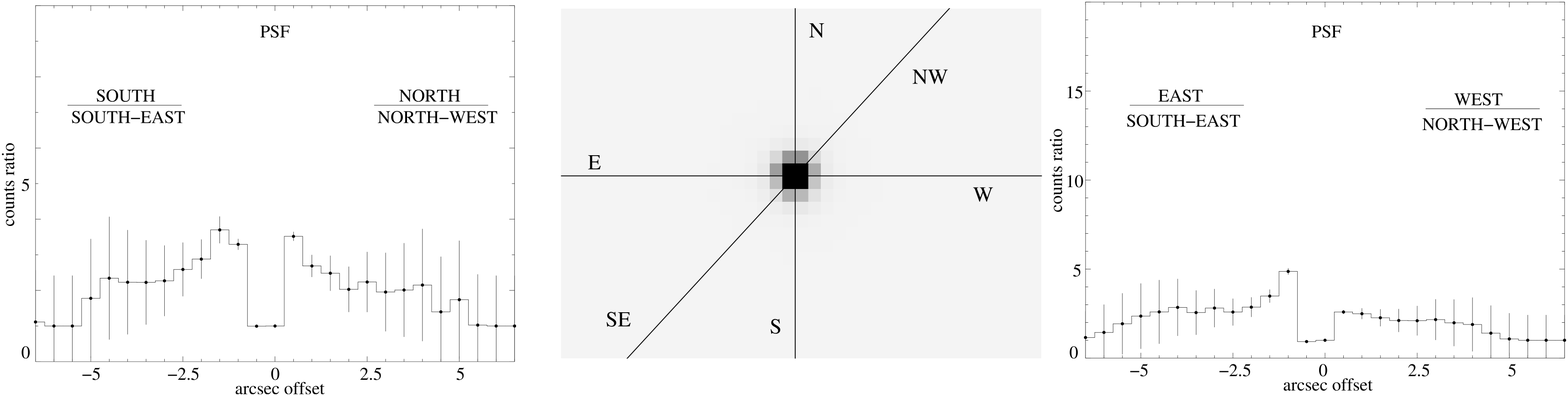,width=15cm,angle=0}\\
\\
\psfig{figure=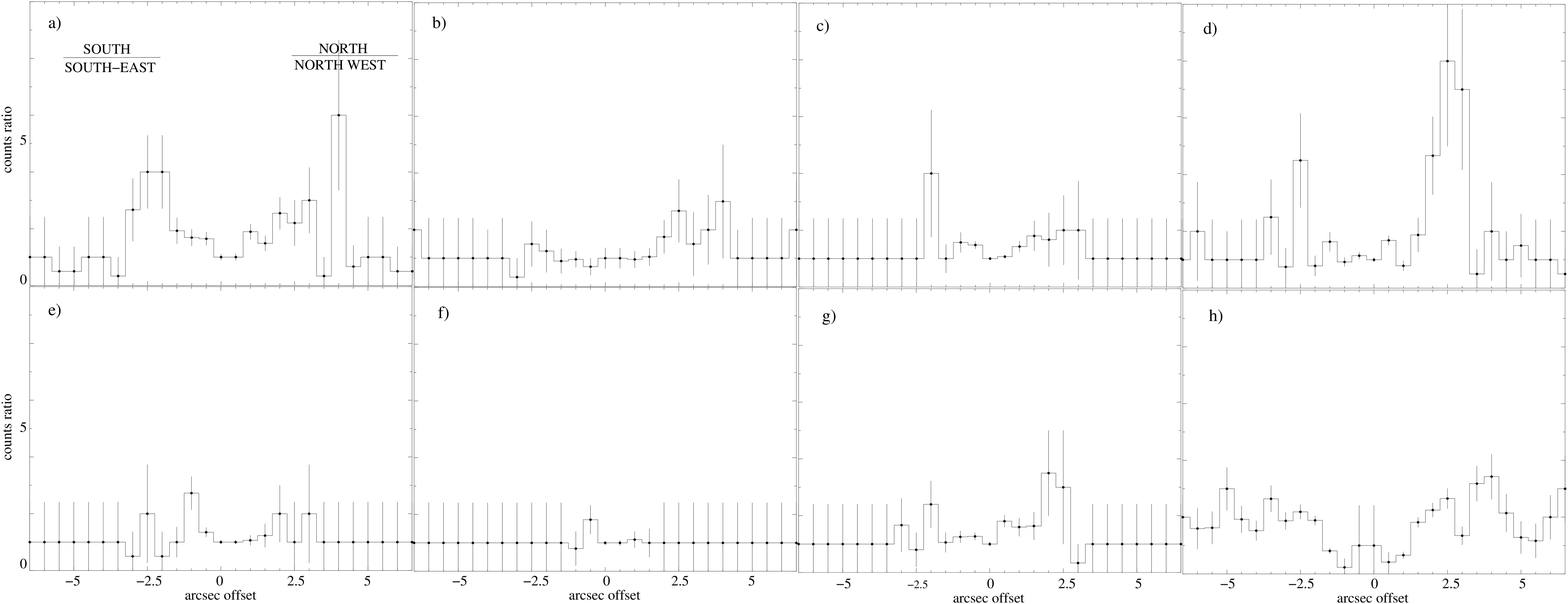,width=18cm,angle=0}\\
\\
\psfig{figure=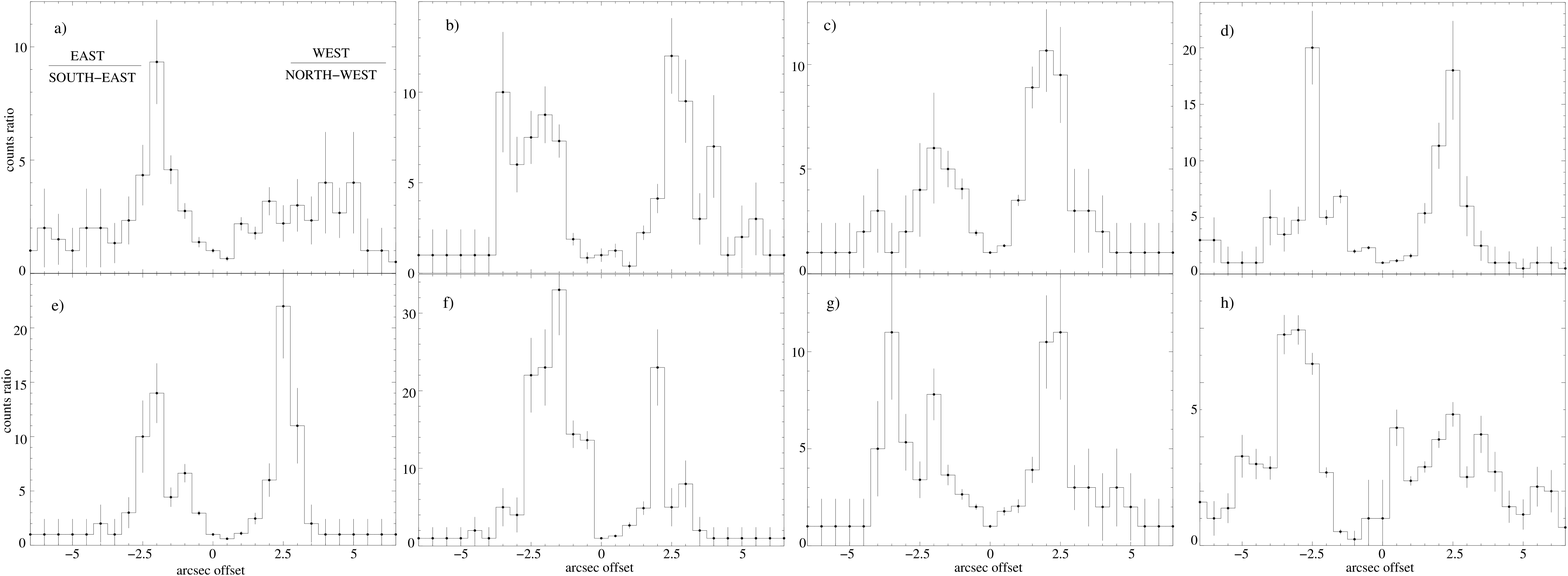,width=18cm,angle=0}\\

\end{tabular}
\caption{{\it Upper panels:} Image of the PSF with the three projections we
have used to analyse the asymmetries in the Chandra images (centre) and
counts ratios of the N-S (left) and E-W (right) by the NW-SE projection as a
function of the distance from the centre of the PSF image. {\it Middle
panels:} Counts ratios of the N-S by the NW-SE projection versus offset from
the centre of the images (see \S~3.1) of all the Chandra images shown in
Fig.~1. All the plots are on the same scale. {\it Lower panels:} Counts ratios
between the E-W and the NW-SE projection versus the offset from the centre of
the images of all the Chandra images shown in Fig.~1. Note the different
scales in the ordinate of the plots.}
\end{figure*}

A simple model was presented in Migliari et al. (2002) to explain the
re-heated arcsec X-ray jets of SS 433 as `average' long-term structures, with
line energies representing the mean Doppler shifts on each side of the
jet. Imaging as presented here with a larger sample of observations suggests
a considerably more complex picture.

\subsection{Morphology}

In order to study the asymmetries of the extended emissions in the X-ray
images shown in Fig.~1, we have analysed the projections of the 0.5-10~keV
images in three directions: north/south (N-S), east/west (E-W) and
north-west/south-east (NW-SE). The angular width of the projections is
$<0.1\arcsec$, i.e. much less than the pixel size of the images. Dividing the
counts in the N-S and E-W projections by the counts in the NW-SE projection,
and comparing these ratios to what expected from the PSFs of the images, we
are able to investigate not only the E-W jet structure, but also the possible
presence of an extended X-ray emission in the equatorial direction (extended
radio emission in a direction approximately perpendicular to the jets has been
observed on milli-arcsec scale; Paragi et al. 1999; Blundell et al. 2001). In
Fig.~4 (central image of the upper panels) we show a typical PSF image and the
ratios of the projections. The PSFs at 3~keV have been extracted with the tool
{\it mkpsf} and normalised to the total counts of the image (Table~1; see
\S~2.1.1 and \S~2.1.2). On either side of the PSF image of Fig.~4 we show the
counts ratios of the N-S/NW-SE and E-W/NW-SE projections versus the offset
from the brightest pixel in the image.  In the middle and lower panels of
Fig.~4 we plot, for each observation, the counts ratios (not background
subtracted) as a function of the offset from the core of the source, showing
the extended asymmetries in the N-S and E-W direction, respectively. The core
position in each image has been chosen as the brightest pixel in the case of
HETGS images, and the pixel in the center of the pile-up distortion in the
case of ACIS-S images. In order to avoid divisions by zero when no counts are
present in the projected pixel, we assigne arbitrarily to the pixel a value of
$1\pm1$ count. Note that, due to precession, the jets are not always perfectly
in the E-W direction and the histograms in Fig.~4 reflect only the component
in this direction. If we compare the plots in the middle panels to the plot of
the PSF in the upper left panel, although there seem to be indications for an
extended emission in a direction perpendicular to the jets in observations
$a$, $d$ and $h$, the asymmetries are consistent within errors with
asymmetries of their PSFs. All the observations show actual extended X-ray
emissions in the E-W direction, not related to asymmetries of the PSF
image. These extended jet emissions have been already shown - and quantified -
for the east jets in Fig.~3 and Table~1.

\subsection{Variability of the X-ray jets}

\subsubsection{Long-term}

Comparison of our two ACIS images (Fig.~1, $b$ and $h$), obtained three years
apart, indicates X-ray structural changes. Including archival Chandra
observations we now have a sample of eight images of the arcsec X-ray jet
structure (Fig.~1). These images (see also Fig.~3) indicate that the arcsec
scale X-ray jets do not have a static and long-term structure. The jet
structure appears, instead, to be continuously evolving. In Table~1 we report
the count rates estimated in the core and in the regions of the east jet
indicated in the upper panel of Fig.~1, for all the Chandra observations.  In
the lower panel of Fig.~3, we plot the count rates from the extended emission
in the east regions after the subtraction of the core contribution (see
\S~2.2). The plot shows significant variations in the X-ray emission from the
east jet. The east jet count rates change significantly between different
observations and even between observations with approximately the same
precession phase (e.g. observations {\em c} and {\em f}), implying no
correlation between the re-heating process and the precession phase of the
jets. In particular in observations {\em e}, {\em f} and {\em g}, taken within
approximately 5 days, basically with the same precession phase, we see very
fast variations of the arcsec structure (see \S~3.2.2) with bright regions
located at different distances from the binary core. This indicates that
re-heating at arcsec scales does not happen at a particular distance from the
binary. Similar results, albeit on smaller physical scales, have recently been
reported by Mioduszewski et al. (2004) based on Very Long Baseline Array
(VLBA) radio observations of the milli-arcsec scale jets of SS~433, in which
they show radio knots getting brighter at different distances from the
core. This is contrary to the previous idea of a fixed (radio) `brightening
zone' around 50 milli-arcsec (Vermeulen et al. 1993a).

\subsubsection{Daily}

\begin{figure}
\centering{\psfig{figure=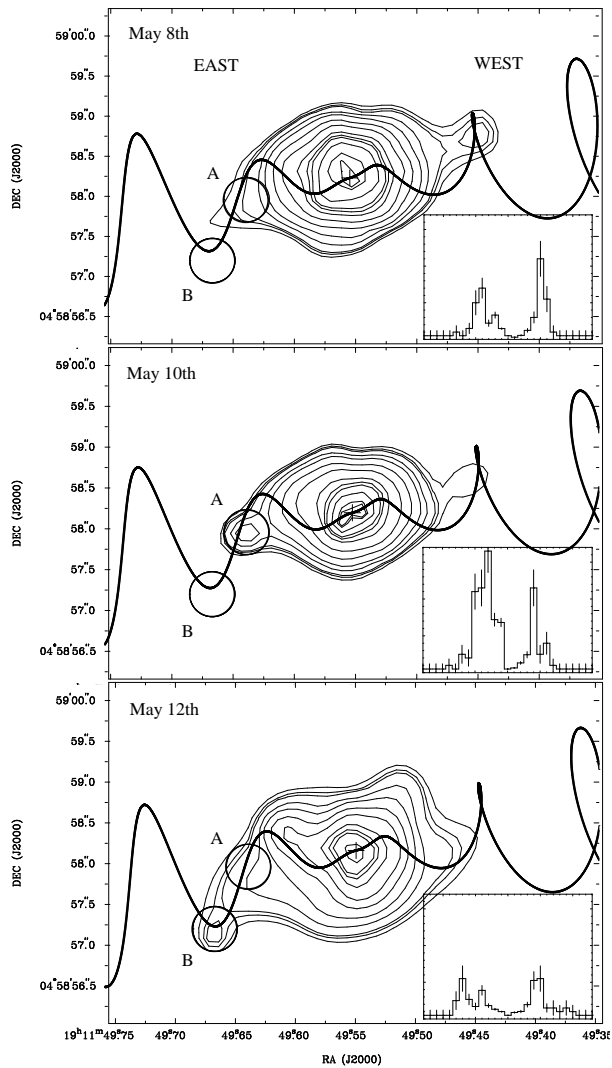,width=8cm,angle=0}}
\caption{Three $\sim20$~ks zeroth order HETGS images of
SS~433, taken every 2 days on 2001, May 8th (top), May 10th (middle), May 12th
(bottom). Images are rebinned to one sixteenth of the original pixel size and
smoothed. Contour levels are 2.7, 3, 3.2, 4, 5, 7, 10, 15, 18, 20, 25, 34, 37
counts per rebinned--pixel for the May 12th observation and normalised to this
for small changes in exposure times (less than $6\%$) in the other 2
observations. The twisted precession trace as predicted from Stirling et
al. (2002), with the small phase correction (see \S~3.2.2), is superimposed to
the images. The lower-right panels show the histograms of the three
observations as in Fig.~4 (but here on the same scale) the variations of the
source in the E-W direction (see \S~3.1). These contour images indicate a
brightening of the east jet and a rapid fading in the west jet. We associate
the X-ray variability with the brightening and fading of unresolved components
projected along the precession trace. The geometry at this epoch is sketched
in Fig.~6.}
\end{figure}
\begin{table}
\centering
\caption{Background subtracted counts, with $1\sigma$ statistical errors, 
in region A and region B (see Fig.~4) for the three 2001 May observations. The
counts are calculated in the 0.5-10~keV energy range and normalised to the
exposure time of the 2001 May 12th observation.}
\begin{tabular}{l l l}
\hline
Obs. & A (counts) & B (counts)\\
\hline
2001 May 8    &$46.1\pm6.8$ & $29.0\pm5.4$\\ 
2001 May 10   &$75.2\pm8.9$ & $38.1\pm6.4$\\ 
2001 May 12   &$62.8\pm7.9$ & $57.8\pm7.6$\\ 
\hline

\end{tabular}
\flushleft

\end{table}
Inspecting three $\sim20$~ks zeroth order HETGS images taken over a period of
$\sim 5$ days ($e$, $f$ and $g$ in Fig.~1), we observe rapid and significant
changes in the arcsec scale X-ray jets on timescales of $\leq 2$ days.  The
structural variations of the jets of these three observations can also be
observed in the lower panels of Fig.~4 ($e$, $f$ and $g$) where the east jet
counts at $\sim-2$~arcsec increase from observation $e$ to observation $f$ and
then decrease again in $g$. Note that since we are observing the counts
projected in the E-W direction, which is not the direction of blob B, we are
following mostly the evolution of blob A.  In Fig.~5 we show the contour plots
of the observations $e$, $f$ and $g$ superimposed to the precession trace of
the jets, shifted in phase of $\Delta\phi\sim-0.1$ (more details are presented
in Blundell \& Bowler 2004; see also Stirling et al. 2004) with respect to the
model in Stirling et al. (2002). (Nevertheless, with or without this small
shift the results of this work are not significantly affected, neither
qualitatively nor quantitatively.) The contour levels correspond to the
smoothed image. The minimum significance signal to noise ratio in the
smoothing procedure has been decreased from 3 (as in Fig.~1) to 2.5 in order
to emphasise the knot structures in the jets. The knot structure is observable
when the counts of the source and the resolution and the sensitivity of the
instrument are high enough (e.g. ACIS-S observations in Fig.~1, $b$ and
Fig.~1, $h$ and the nearly 30 ks HETGS observation in Fig.~1, $a$). We have
chosen the size of the circular regions in which we estimate the counts, as
large as the original-size pixel of the image, and such that outside the
circle regions the counts per rebinned-pixel of a point source distributed on
the image by the PSF are less than the 25 per cent of the counts per
rebinned-pixel at the peak.  We observe a knot getting brighter from May 8th
to May 10th (knot A) and subsequently on May 12th we see it apparently moving,
following the precession trace (knot B). This indicates that we are observing
two knots lying on the precession trace, at different precession phases, which
brighten up sequentially. To quantify these brightenings we have estimated the
background subtracted counts in the circular regions A and B for these three
observations (see Table 3). The counts have been normalised to the exposure
time of the May 12th observation. Taking the counts in the regions on May 8th
as a reference, we observe region A brightening $\sim$two days later and
subsequently region B getting brighter, $\sim$four days later. Considering
statistical errors, both brightening are significant at the $>99\%$ confidence
level if we consider May 8 knots as a reference: knot A: May 10$-$May
8=$29.1\pm11.2$, i.e. 99.1\% significance; knot B: May 12$-$May
8=$28.8\pm9.3$, i.e. 99.7\% significance. (When testing against the null
hypothesis of a constant model, the probability of obtaining a larger value of
$\chi^2$ is 3\% for knot A and 1\% for knot B.) Since {\em i)} the core count
rate decreases from May 8th to May 10th while the count rate of knot A
increases and {\em ii)} we observe a (albeit small) decrease in the count rate
of knot A (which is closer to the core than knot B) from May 10th to May 12th,
while the count rate of knot B increases, and since the PSF of the core is
about symmetric at those distances from the core, we are confident that we are
observing actual variations in the jets, not directly related to
contaminations of the PSF of the core.

\subsubsection{Hour-timescale}

Analysing the 60 ks ACIS-S observation of SS 433 (Fig.~1, $h$) we have
selected two regions of the image corresponding to the east (the region is
smaller than the one shown in Fig.~1, $h$ and zoomed on the two knots clearly
visible in the image) and the west jets. We have applied baricenter
corrections and extracted the background subtracted light curves in these
regions with a time bin of $\sim6000$~s, in the range 0.5-10~keV. We have fit
the two light curves with a constant and did a $\chi^{2}$ test to check if
they are consistent with a steady jet emission on time scales of hours. The
west jet is consistent with being steady. However, in the light curve of the
east jet, although it does not show any systematic trend, we can reject a
steady distribution at the 3.5 per cent level (the fit with a constant gives a
$reduced-\chi^2=2.0$ with 9~$d.o.f.$). Assuming a variation in the east jet -
the variation is marginally significant and needs to be confirmed - because of
the signal propagation speed, the linear size of the emitting region should be
smaller than $c\times\Delta t \sim 10s$ AU, where $\Delta t\sim60ks$. We have
also extracted the lightcurve, with the same time bin, of the readout streak
of the image, to check the variability of the core. The core is consistent
with being steady over the whole observation.

\section{Jets re-heating: possible scenarios}

Whatever the underlying explanation, the rapid variability indicates that the
observed extended X-ray emission originates in physically small components
which may be suffering shock acceleration, as we now consider. The lack of
correlation with precession phase indicates an origin of this phenomenon in a
more random process, e.g. the variability of the underlying accretion flow.

\subsection{Internal shocks}

The possibility of internal shocks (orginally proposed by Rees 1978 for the
radio galaxy M87) between `blobs' ejected at almost the same precession phase
with slightly different velocities (the mean speed of the thermal matter in
the jet is $\sim0.26c$ and might be scattered of $\pm15$ per cent: Eikenberry
et al. 2001; see also Blundell and Bowler 2004) can explain the arcsec-scale jet X-ray brightenings (Migliari et
al. 2002). Knot brightenings similar to those described above in X-ray are
also observed in the radio band on much smaller physical scales
(e.g. Vermeulen et al. 1993a). Mioduszewski et al. (2004) have recently
monitored with the VLBA the milli-arcsec scale variability of SS~433 for
one-fourth of the jet precession cycle. In the 42 days of their radio
monitoring we observe an average of a blob ejection every $\sim6$ days, with a
peak of one every 2-3 days in the most active period. In 6 days the jet axis
has moved, due to precession, of about $1.5^{\circ}$. With such a small angle,
two $\sim5^{\circ}$-size blobs (e.g. Begelman et al. 1980) can still interact
with each other if the second blob has been launched faster than the first,
although they move in slightly different directions. In the case
of a blob with a velocity of $0.3c$ launched 6 days after a blob with a
velocity of $0.2c$, the second blob should reach the first at a distance of
$\sim10^{16}$~cm from the core. To explain knots getting brighter at distances
of $\sim10^{17}$~cm from the core, where we observe the arcsec-scale X-ray
jets, the difference in the blobs' velocities can be smaller and an `internal
shock' mechanism may explain what we observe in X-rays.

If this is the correct scenario and since we observe in both X-ray (Fig.~1)
and radio images (Mioduszewski et al. 2004) knots getting brighter with no
dependence on distance or precession phase, the velocity scattering of the
ejections should be more or less random with respect to periodicities of the
system.  Therefore, we must note that the fact that the first time the arcsec
X-ray jets of SS~433 have been `monitored' with three observations within
about five days (observations $e$, $f$, $g$) show what seems to be a
`sequence' of events where two knots get brighter one after the other (knot A,
and then knot B in Fig.~5), is at least curious and worth
investigating. Moreover, Mioduszewski et al. (2004) show that radio knots can
get brighter at distances of $\sim10^{15}$~cm from the core. This distance
seems too short for an internal shock scenario which considers two blobs
launched 6~days (but even 2-3 days) apart with a velocity difference of only
$\sim15$ per cent (i.e. a factor of more than 2 in the speed of one of the
blobs is still needed). Based on this, we suggest that something else might be
at the origin of the knot brightening in the jets, which can account for both
large and small-scale observations and for the sequence of brightenings
observed in X-rays. We suggest a new scenario in which the slower-moving
clouds are energised erratically by a more powerful, faster, unseen flow.

\subsection{Proposed scenario: an underlying faster outflow}

\begin{figure}
\centering{\psfig{figure=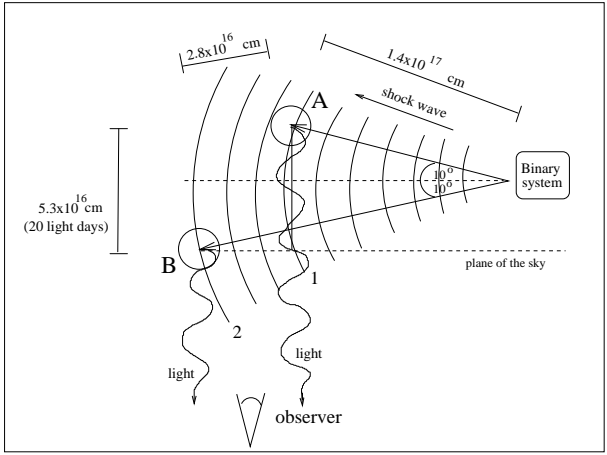,width=8cm,angle=0}}
\caption{Sketch of the orientation of the east jet of SS
  433 at the time of the observations in Fig.~5 (assuming a distance of 5
  kpc). Knot B was observed to brighten $< 2$ days after knot A.  We propose
  that a shock wave propagates from the binary core (on the right) through the
  jet and hits first the knot A (1) and then the knot B (2). Knot A is about
  20 light days farther from us than knot B. We see knot B getting bright at
  least 2 days later than knot A. This means that the shock wave has to travel
  from knot A to knot B within about 22 days, with a velocity of $\ga0.5c$.}

\end{figure}

In Fig.~5 we observe two knots, corresponding to regions A and B. The
kinematic model tells us the projection onto the plane of the sky of the knots
ejected. By identifying these knots on the precession trace we can de-project
the image and give a three dimensional position of the knots. A sketched view
from ``above'' of the east jet as observed in Fig.~5 is shown in Fig.~6. As
described in \S~4.1, we assume that a shock wave forms and travels at
relativistic velocities (faster than the knots' velocity which is about
0.26~c) through the jet. Assuming a distance to the source of 5~kpc, we can now
estimate the time scales and spatial scales involved in such shock events . In
the May 10th image of Fig.~5 we see that knot A is about in the middle of the
`descending' trace at a distance of about $1.4\times10^{17}$~cm from the
binary core. Following the kinematic model, we infer that knot A was ejected
about 200 days before the observation, and is moving away from us at an angle
to the line of sight of $\sim 100^{\circ}$ (see Fig.~6). The May 12th image in
Fig.~5 shows knot B at the bottom of the precession trace at a distance of
about $1.7\times10^{17}$~cm from the core. Knot B was ejected from the binary
system about 40 days earlier than knot A, and is moving towards us with an
angle to the line of sight of $\sim 80^{\circ}$ (see Fig.~6).  The shock wave
interacts first with knot A and then with knot B. At the time at which the
shock interacts with knot A, this was about 20 light days farther from us than
knot B. Therefore, the shock wave has to travel the projected distance between
the two knots in about 22 days (we observe knot B brightening no more than two
days after knot A). In order to observe the brightenings of the two components
within two days, the shock has to travel with a velocity of $v\sim 0.5c$ [the
velocity is $\sim0.6c$ if we use the most recent estimate for the distance of
5.5~kpc (Blundell \& Bowler 2004)]. Note that the value we have inferred of
$0.5c$ might be considered as a lower limit because i) the positions of the
knots on the precession trace were chosen by eye from Fig.~5, any other
position on the `descending' trace of any of the two knots would decrease the
light days difference to us between the knots and thus increase the velocity
of the shock; ii) the brightening of knot B after knot A might have happened
anytime within the 2 days between May 10th and May 12th observations, thus,
since usually the radiative lifetime of the thermal knots is of the order of
two days or more (Vermeulen et al. 1993b), we may consider 2 days as an upper
limit in time. What we observe is therefore consistent with being due to a
shock wave which travels from the binary core downstream in the jet with an
inferred velocity of $\ga0.5c$. This shock should be broader than the thermal
jet collimation angle of $\sim5^{\circ}$ as inferred by optical spectral
analysis at distances of $10^{15}$~cm from the core, because it has to
brighten up two knots with different precession phase (i.e. angle to the line
of sight). Since the velocity of this shock wave is greater than that observed
for the moving components (including the possible scatter of 15 per cent;
Eikenberry et al. 2001), it seems to correspond to an unseen second component
to the outflow.

\section{Discussion}

Eight X-ray Chandra images of SS~433 revealed that its arcsec X-ray jet
structure varies very rapidly, possibly as short as hour timescales. The
observations show that the reheating processes, which are at the origin of the
X-ray jets, have no preference for a particular precession phase of the jet or
distance from the binary core. In particular, a sequence of three images made
within about five days show two knots in the east jet getting brighter
sequentially, suggesting that both knots are responding to the same
phenomenon.

Radio changes, like those described above in X-rays, have been observed also
on smaller scales (of the order of $10^{16}$~cm from the binary core;
Vermeulen et al. 1993a; see also Mioduszewski et al. 2004). X-ray images
indicate that the shocks which may cause those changes in radio continue also
on larger scales. Moreover, since the knots fade significantly on timescales
of days (bullets which emit optical line have a typical lifetime of $\sim2$
days: Vermeulen et al. 1993b; Fig.~5 also shows a knot in the west jet fading
in a few days), and assuming that the brightening of the two knots A and B are
responding to the same underlying phenomenon, a frequent `pumping' of such
shocks is required. Put in another way, if the core energy source switches
off, we would expect the arcsec-scale X-ray jets to fade away very rapidly.

Another possibility, other than a single shock wave, is sequence of shocks,
like a wave train. In this case there is the possibility that knot A and knot
B have been brightened by two different crests of the same wave train. If on
May 8th the first crest has already passed knot A, it has to be since about
2-6 days, which is the range in which we expect the knot to fade away almost
completely. Therefore, if this first crest brightens knot B on May 12th, it
takes $\sim8$~days to travel $2.8\times10^{16}$~cm (see Fig.~6), implying a
slightly slower shock wave with a velocity of $\sim0.4c$. If a second crest of
the wave train brightens knot A on May 10th, the spacing between the crests is
of a few days.

The possible existence of an underlying, energising flow which is not directly
observed and has a velocity greater than the observed components might have
seemed speculative only a few years ago. However, recent detailed studies of
the jets from two neutron star X-ray binaries, Sco X-1 (Fomalont et al. 2001)
and Cir X-1 (Fender et al. 2004), have revealed precisely this phenomenon. In
Sco X-1 Fomalont et al. (2001) observed compact radio jet lobes moving outward
with a velocity of at most $0.57c$. They observed that flares in the core and
in the lobes are correlated and indicate an energy flow moving from the core
within the jet beams with a speed of $>0.95c$. A similar behaviour has been
observed in Cir~X-1. Fender et al. (2004) observed an ultrarelativistic energy
flux moving from the core with a velocity of $>0.99c$ and brightening
consequently two radio jet lobes of the approaching jet which are observed to
move much slower. The two lobes which get brighter in Cir~X-1 are not
precisely aligned with the core (although a very small angle of the jet to the
line of sight, as Cir~X-1 seems to have, would amplify the apparent
misalignement), further supporting the idea that the second component
relativistic jet flow might be less collimated than the slower discrete knot
ejections observed getting brighter. The observations presented here challenge
our current understanding of SS~433, which has been considered one of the best
known galactic source to date.

\section*{Acknowledgements}

We thank the referee, Herman Marshall, for insightful comments which helped
to improve the paper. SM wishes to thank Gabriele Ghisellini for stimulating
discussions.

\end{document}